\documentclass{article}
\usepackage{graphicx}

\def\abstract#1{\vskip 7mm
        \begin{center}{\large Abstract}\par \smallskip
                \begin{minipage}[c]{12cm}
                        \small #1
                \end{minipage}
        \end{center}
}
\def\title#1{\begin{center}{\Large\bf #1}\end{center}}
\def\author#1{\vskip 5mm \begin{center}{#1}\end{center}}
\def\address#1{\begin{center}{\it #1}\end{center}}
\makeatletter
\@ifundefined{lesssim}{}{}
\@ifundefined{gtrsim}{}{}
\def\vereq#1#2{\lower3pt\vbox{\baselineskip1.5pt \lineskip1.5pt
\ialign{$\m@th#1\hfill##\hfil$\crcr#2\crcr\sim\crcr}}}
\makeatother

\def\black{}
\def\red{}

\def\blue{}

\def\magenta{}

\begin{document}

\title{The disinformation problem for black holes}

\author{\blue Sean A. Hayward\black}

\address{Department of Physics, National Central University, Jhongli,
Taoyuan 320, Taiwan}

\noindent\magenta{\em No escape?}\black

\medskip\noindent1966: ``{\em event horizon}\dots is the boundary of the region from
which particles or photons can escape to infinity\dots a {\em black hole} is a
region\dots from which particles or photons cannot escape''

\medskip\noindent1976: ``Because part of the information about the state of the system
is lost down the hole, the final situation is represented by a density matrix
rather than a pure quantum state''

\medskip\noindent1997: ``Whereas Stephen Hawking and Kip Thorne firmly believe that
information swallowed by a black hole is forever hidden from the outside
universe, and can never be revealed even as the black hole evaporates and
completely disappears\dots''

\medskip\noindent2004: ``Thus the total path integral is unitary and information is
not lost in the formation and evaporation of black holes. The way the
information gets out seems to be that a true event horizon never forms, just an
apparent horizon''\hfill[Hawking quotes]

\bigskip\noindent\blue How should black holes be defined? \black Event and
apparent horizons are too indirectly defined.

\noindent Assume spherical symmetry for simplicity (everything generalizes).

\noindent Area \red$A\black$, null coordinates \red$x^\pm\black$:
$g^{-1}(dx^\pm,dx^\pm)=0$, future-pointing, unique up to $x^\pm\mapsto\tilde
x^\pm(x^\pm)$.

\noindent Null expansions \red$\theta_\pm\black=\partial_\pm A/A$,
$\partial_\pm=\partial/\partial x^\pm$.

\noindent Normally, outgoing light rays diverge, $\theta_+>0$,

ingoing light rays converge, $\theta_-<0$;

outgoing wavefront expands,

ingoing wavefront contracts:

\vskip-2cm\hfill\includegraphics[height=25mm]{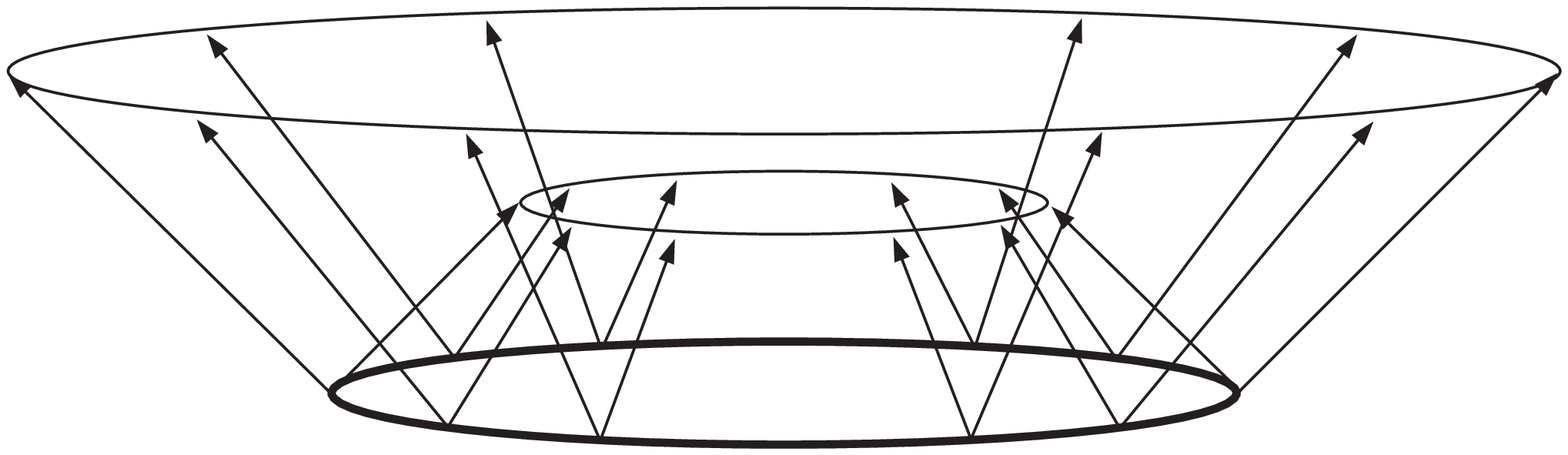}

\noindent But if there is a massive source inside the surface,

its gravitational field has an attractive or converging effect.

Close enough to a massive enough source,

the outgoing light rays may also be converging, $\theta_+<0$; a \red trapped
surface{}\black:

\medskip Everything inside is trapped within a shrinking area.

Nothing can escape, not even light.

\vskip-25mm\hfill\includegraphics[height=25mm]{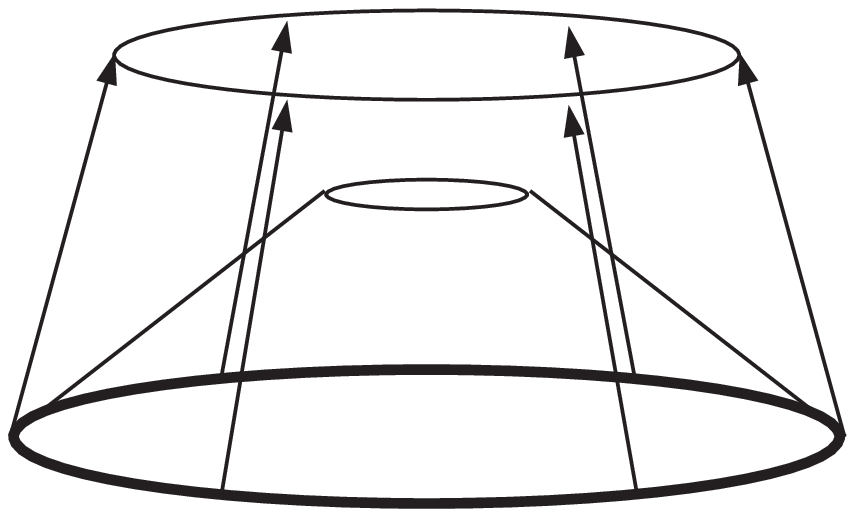}

\noindent In between, there will be a \red marginal surface \black (marginally
trapped surface), $\theta_+=0$,

where the outgoing light rays are instantaneously parallel:

\hfill\includegraphics[height=25mm]{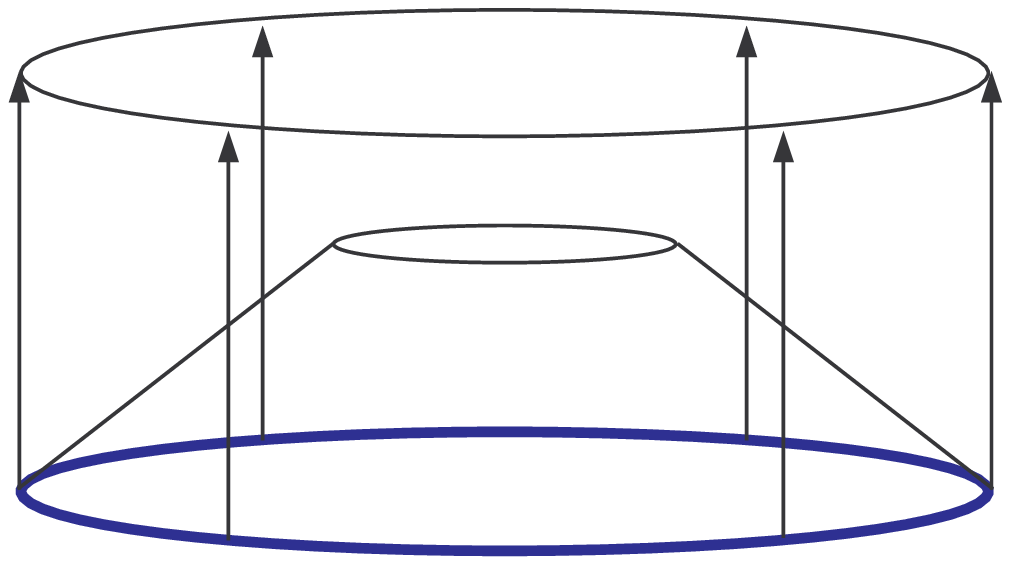}

\vskip-2cm\includegraphics[height=3cm]{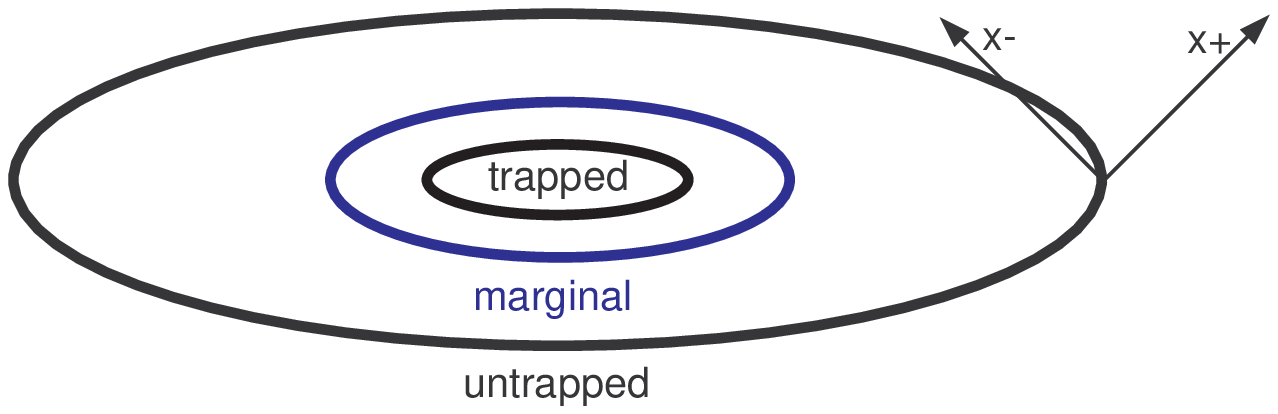}

\noindent This is a black hole: its surface is located by the marginal surface,
where outgoing light rays are instantaneously parallel, $\theta_+=0$, ingoing
light rays are converging, $\theta_-<0$, and outgoing light rays are diverging
just outside and converging just inside, $\partial_-\theta_+<0$. As time
develops, the marginal surfaces generate a hypersurface in space-time.

\vfill\eject\noindent\magenta{\em Gravity traps}\black

\vskip-5mm
$$\hbox{A surface
is}\left\{\matrix{\red\hbox{untrapped}\cr\hbox{marginal}\cr\hbox{trapped}\black}\right\}\hbox{if
$\theta_+\theta_-$}\left\{\matrix{<0\cr=0\cr>0}\right\},$$

$$\left\{\matrix{\red\hbox{future}\cr\hbox{past}\black}\right\}\hbox{trapped if $\theta_\pm$}
\left\{\matrix{<0\cr>0}\right\},
\left\{\matrix{\hbox{future}\cr\hbox{past}}\right\}\hbox{marginal if
$\theta_-$} \left\{\matrix{<0\cr>0}\right\}\hbox{(for $\theta_+=0$).}$$

\noindent\red Trapping horizon{}\black: a hypersurface foliated by marginal
surfaces,
$$\left\{\matrix{\red\hbox{outer}\cr\hbox{degenerate}\cr\hbox{inner}\black}\right\}\hbox{if
$\partial_-\theta_+$}\left\{\matrix{<0\cr=0\cr>0}\right\}\hbox{(for
$\theta_+=0$).}$$

\vskip-5mm
$$\hbox{A}\left\{\matrix{\hbox{future}\cr\hbox{past}\black}\right\}
\hbox{outer trapping horizon can be taken as the local definition of
a}\left\{\matrix{\red\hbox{black}\cr\hbox{white}\black}\right\}\red\hbox{hole{}\black.}$$

\noindent e.g.\ Schwarzschild and Reissner-Nordstr\"om black holes:

\hfill\includegraphics[height=4cm]{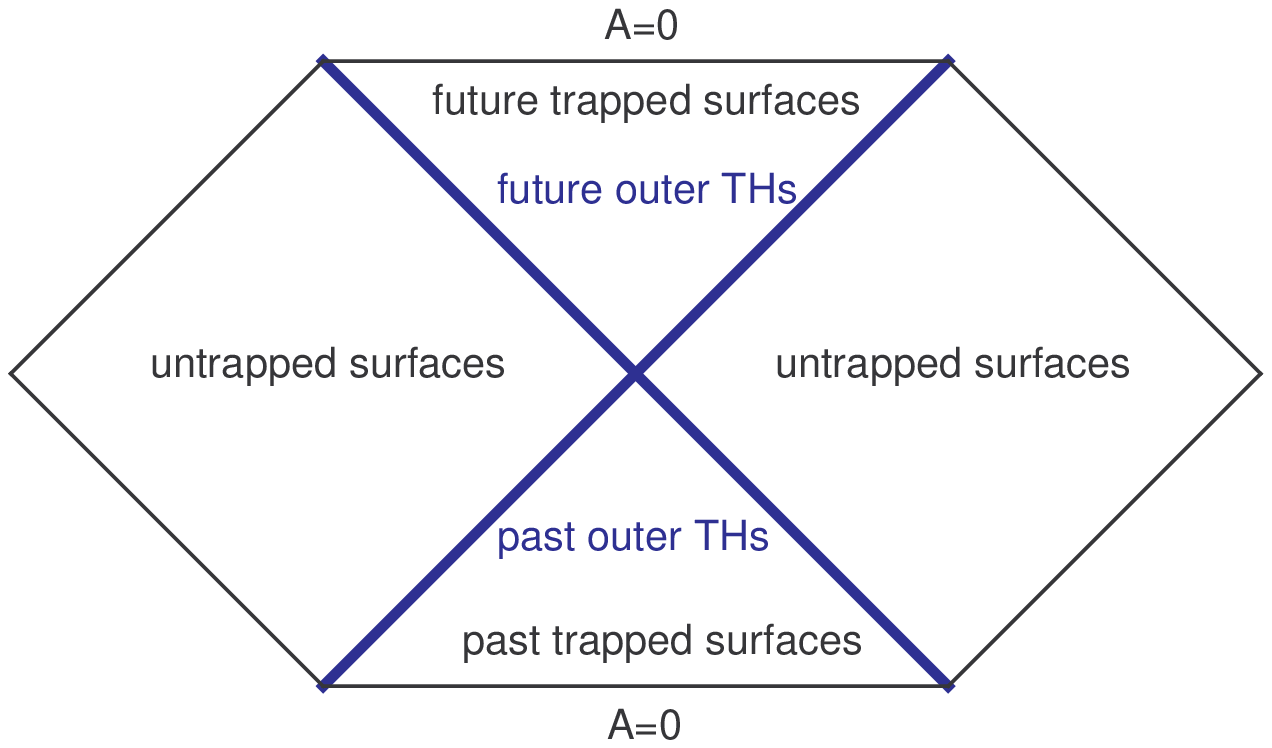}\includegraphics[height=6cm]{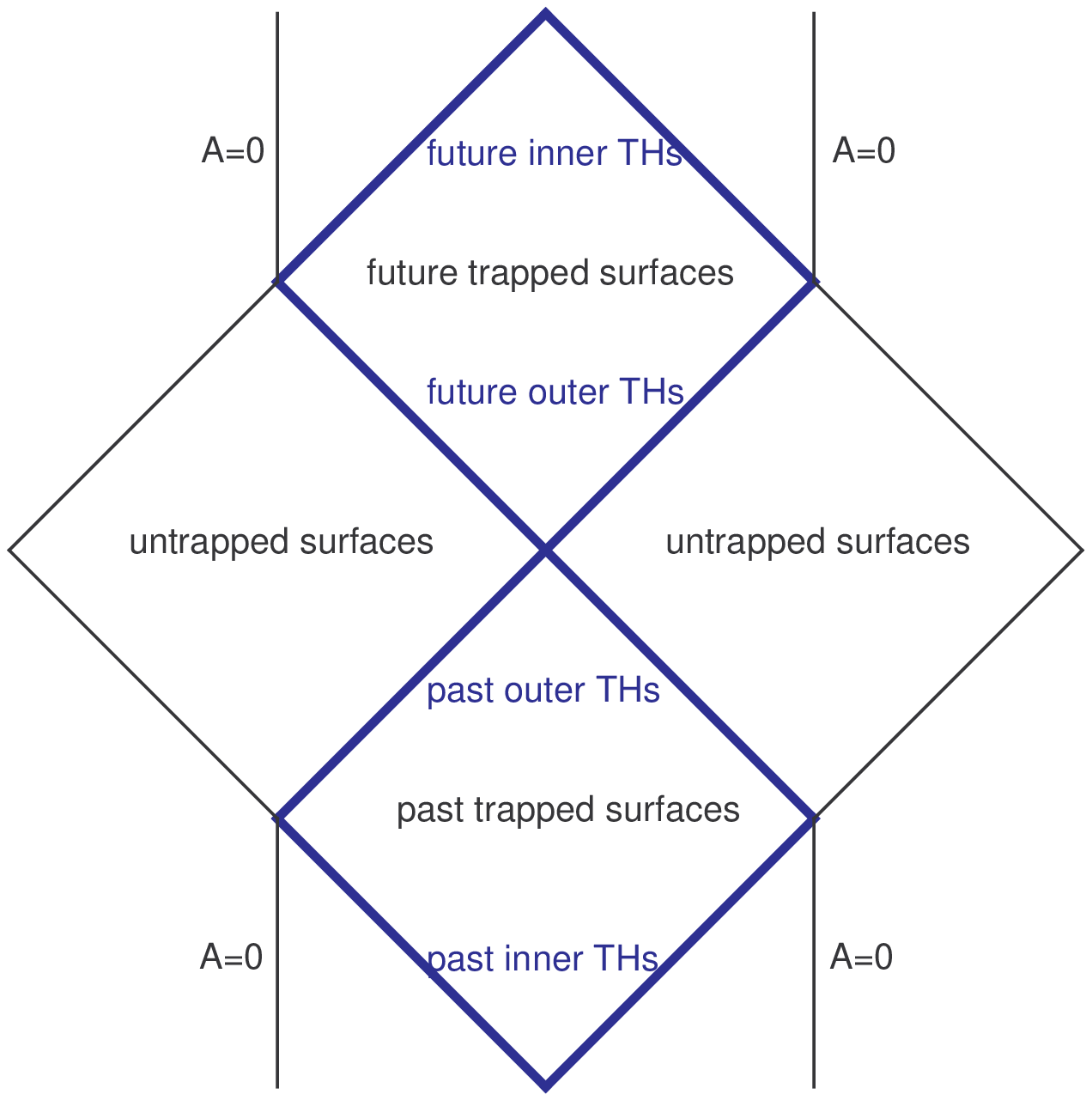}

\noindent\red Signature law\black{}:
$$\hbox{NEC (null energy condition)}\Rightarrow\left\{\matrix{\hbox{outer}\cr\hbox{inner}}\right\}
\hbox{trapping horizons are}\left\{\matrix{\hbox{spatial or
null}\cr\hbox{temporal or null}}\right\},$$

$$\hbox{and null if and only if the ingoing energy density vanishes.}$$

\noindent Proof. Take the case $\theta_+\cong0$ (evaluation on the horizon),

write the horizon-generating vector $\xi=\xi^+\partial_++\xi^-\partial_-$,

then $0\cong\xi\cdot d\theta_+=\xi^+\partial_+\theta_++\xi^-\partial_-\theta_+$
relates three signs:

\vskip-5mm
$$\frac{\xi^-}{\xi^+}\left\{\matrix{<0\cr=0\cr>0}\right\}\hbox{for a}
\left\{\matrix{\hbox{spatial}\cr\hbox{null}\cr\hbox{temporal}}\right\}\hbox{trapping
horizon;}\qquad\qquad\qquad\qquad\qquad\qquad\qquad\qquad\qquad\qquad\qquad{}$$

\vskip-5mm
$$\partial_-\theta_+\left\{\matrix{<0\cr>0}\right\}\hbox{for an}
\left\{\matrix{\hbox{outer}\cr\hbox{inner}}\right\}\hbox{trapping
horizon;}\qquad\qquad\qquad\qquad\qquad\qquad\qquad\qquad\qquad\qquad\qquad{}$$

\vskip-35mm\hfill\includegraphics[height=35mm]{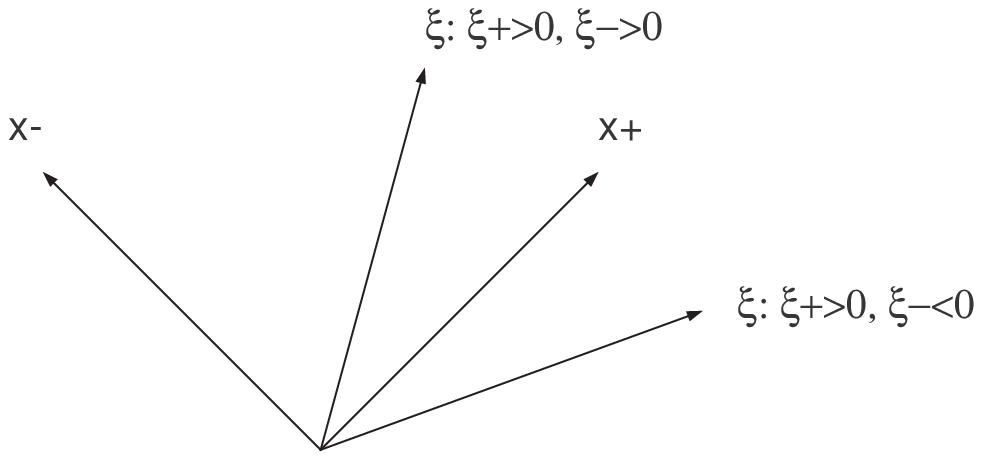}

$\partial_+\theta_+\cong-8\pi GT_{++}$ (null focusing equation, from Einstein
equation),

$T_{\pm\pm}=T(\partial_\pm,\partial_\pm)$ (energy tensor), NEC $\Rightarrow
T_{\pm\pm}\ge0$.

\hfill\includegraphics[height=25mm]{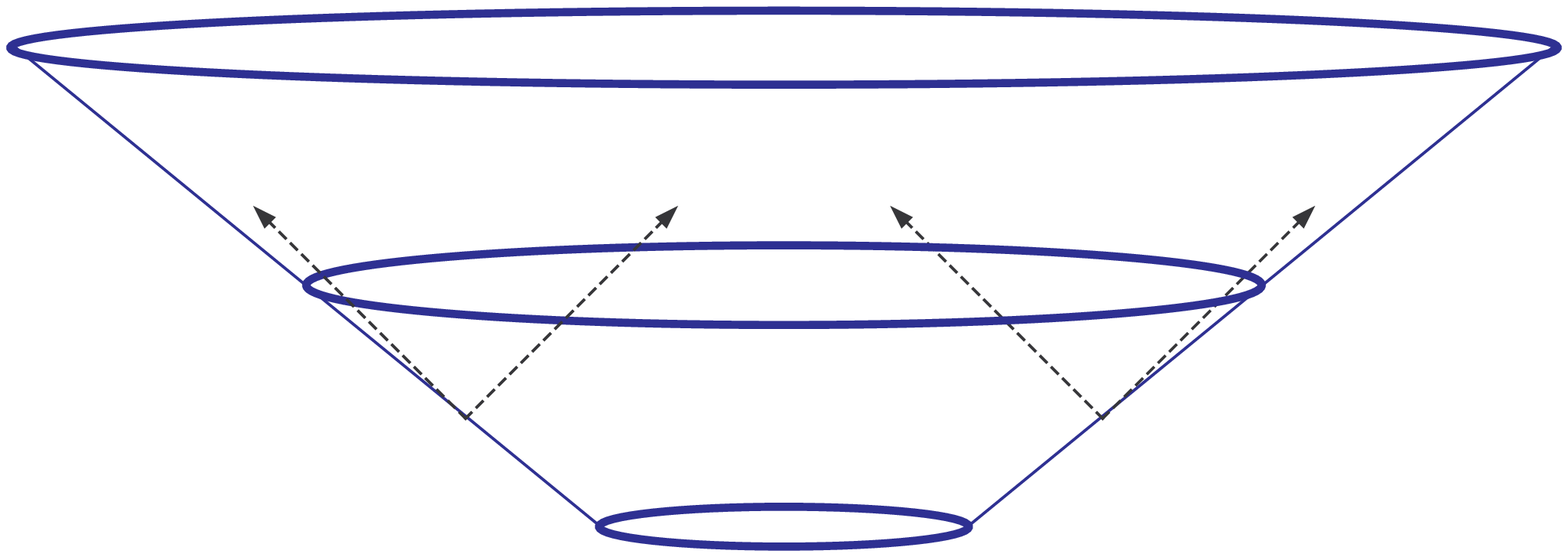}

\vskip-2cm\noindent$\Rightarrow\blue$ Black hole horizons are one-way
traversable\black{};

one can fall into a black hole but not escape:

\hfill\eject\noindent\red Area law\black{}: $\hbox{NEC}\Rightarrow$
$$\left\{\matrix{\hbox{future outer or past inner}\cr\hbox{past outer or future inner}}\right\}
\hbox{trapping horizons
have}\left\{\matrix{\hbox{non-decreasing}\cr\hbox{non-increasing}}\right\}\hbox{area},
\left\{\matrix{A'\ge0\cr A'\le0}\right\},$$

$$\hbox{instantaneously constant ($A'=0$) if and only if the horizon is
null.}$$

\noindent Proof. $A'=\xi\cdot dA=A(\xi^+\theta_++\xi^-\theta_-)\cong
A\xi^-\theta_-$. Fixing orientation $\xi^+>0$,

$$\hbox{$\xi^-$}\left\{\matrix{\le0\quad\hbox{for
outer horizons}\cr \ge0\quad\hbox{for inner horizons}\cr=0\quad\hbox{for null
horizons}}\right\}\hbox{as above; }
\theta_-\left\{\matrix{<0\cr>0}\right\}\hbox{for}\left\{\matrix{\hbox{future}\cr\hbox{past}\black}\right\}\hbox{
horizons}.$$

\medskip\noindent$\Rightarrow\blue$ Black holes grow \black if they absorb any
matter, and otherwise remain the same size.

\medskip\noindent(Actually general results, not restricted to spherical
symmetry).

\medskip\noindent\blue Evaporating black holes\black{}: Hawking radiation is based on
pair production of

positive-energy outgoing radiation, $T_{--}>0$,

and negative-energy ingoing radiation, $T_{++}<0$.

$\Rightarrow$ area and signature properties reverse:

$$\hbox{ingoing energy density negative}\Rightarrow\left\{\matrix{\hbox{outer}\cr\hbox{inner}}\right\}
\hbox{trapping horizons
are}\left\{\matrix{\hbox{temporal}\cr\hbox{spatial}}\right\},\qquad\qquad\qquad$$

$$\left\{\matrix{\hbox{future outer or past inner}\cr\hbox{future inner or past outer}}\right\}
\hbox{trapping horizons
have}\left\{\matrix{\hbox{decreasing}\cr\hbox{increasing}}\right\}\hbox{area},
\left\{\matrix{A'<0\cr A'>0}\right\}.$$

\bigskip\noindent\blue The black-hole horizon is shrinking and two-way traversable\black{}:

Matter and information clearly will escape.

\vskip-1cm\hfill\includegraphics[height=3cm]{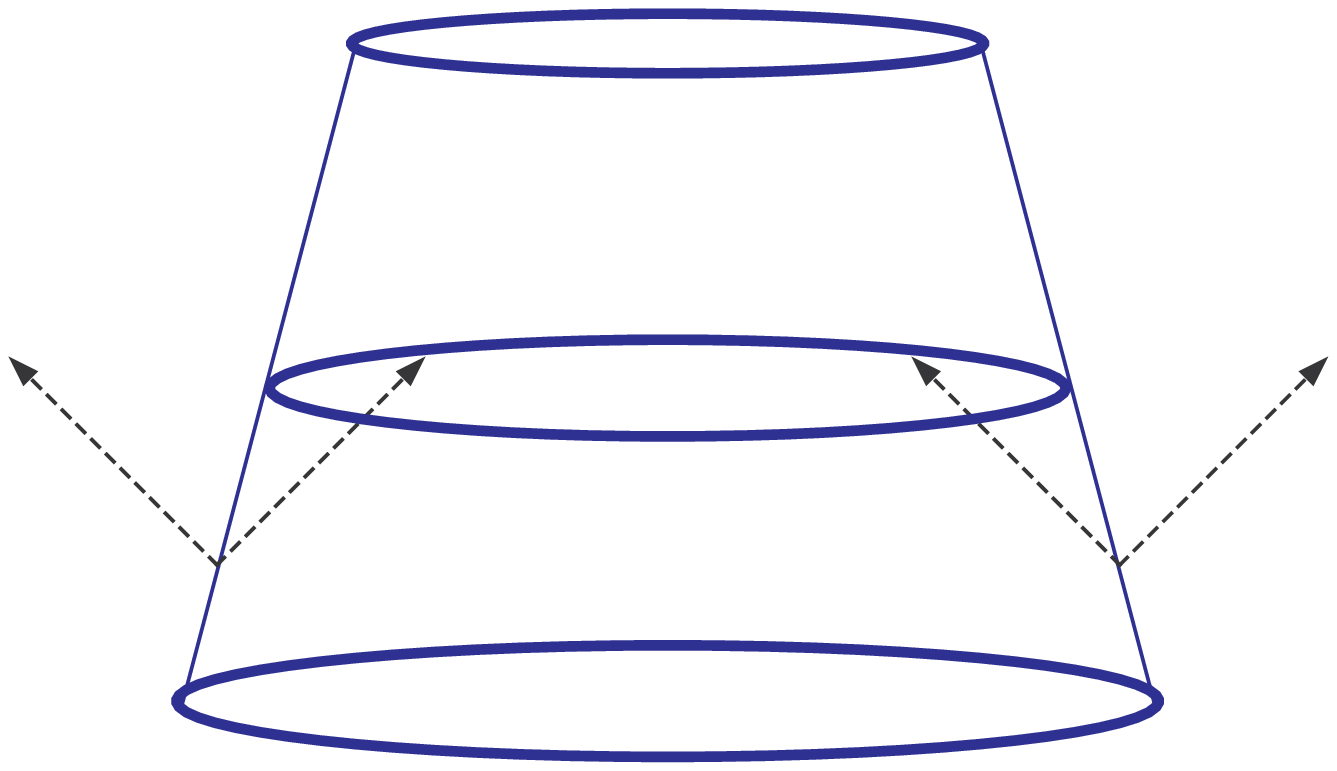}

\noindent\magenta{\em Losing it?}\black

\noindent No reason to expect a purely thermal spectrum for an evaporating
black hole;

Hawking's thermal spectrum holds only for a stationary black hole,

ignoring the back-reaction of the radiation on the black hole, which is not
evaporating.

\bigskip\noindent\blue Endpoint of evaporation? \black Usual picture:

\vskip-1cm\hfill\includegraphics[height=4cm]{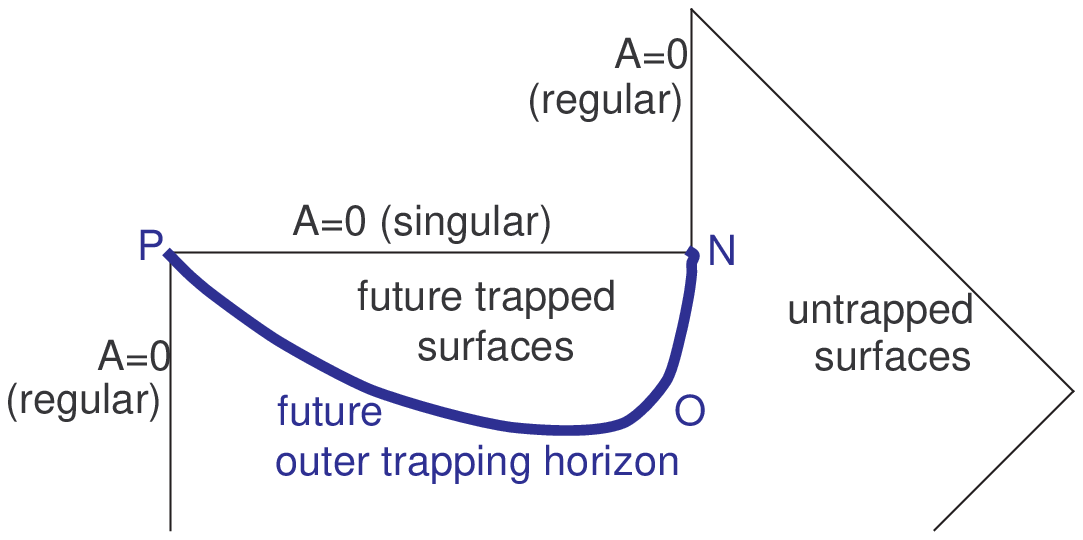}\vskip-1cm

PO: $T_{++}>0$, outer $\Rightarrow$ spatial, growing.

ON: $T_{++}<0$, outer $\Rightarrow$ temporal, shrinking.

The horizon shrinks to zero size, intersecting the singularity.

Some of the matter (and information) escapes, some is lost in the singularity.

However, this is a semi-classical approximation, valid only away from the
singularity.

In full quantum gravity, is the singularity resolved?

\vfill\eject\hfill\includegraphics[height=6cm]{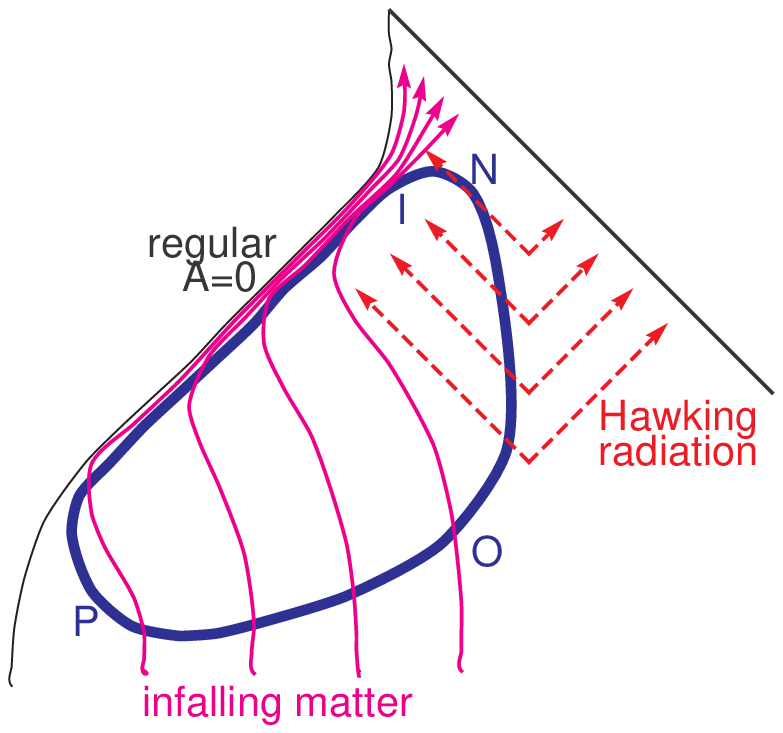}

\vskip-6cm\noindent\magenta{\em The great escape}\black

\noindent Assume that the centre never becomes singular.

Regular centre $\Rightarrow\exists$ untrapped neighbourhood

$\Rightarrow\exists$ inner (future) trapping horizon.

\noindent Inner horizon and centre approach, become almost null,

matter becomes almost pure outgoing radiation:

($\exists$ exact Einstein-Klein-Gordon solution).

\noindent During black-hole formation:

PO: $T_{++}>0$, outer $\Rightarrow$ spatial, growing.

PI: $T_{++}>0$, inner $\Rightarrow$ temporal, shrinking.

\noindent During black-hole evaporation:

ON: $T_{++}<0$, outer $\Rightarrow$ temporal, shrinking.

IN: $T_{++}<0$, inner $\Rightarrow$ spatial, growing.

The outer and inner sections eventually meet,

marking the endpoint of evaporation.

\noindent Metric $C^2\Rightarrow$ inner and outer sections join smoothly.

\noindent $\Rightarrow\exists$ \blue single trapping horizon of $S^2\times S^1$
topology\black{},

enclosing a compact region of trapped surfaces.

\medskip\noindent The black hole has evaporated completely,

with all its contents re-emerging, albeit mangled.

\vskip-3cm\hfill\includegraphics[height=4cm]{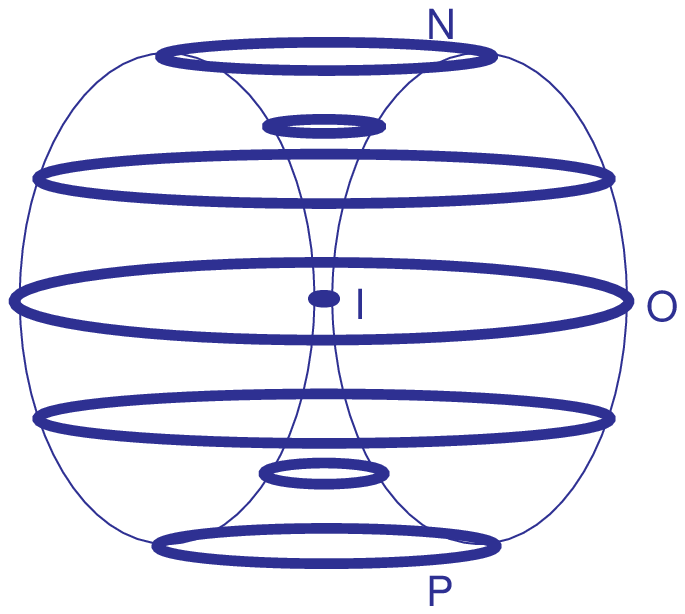}

\noindent\blue No information loss; no event horizon\black{}. Obtained by (at
least) three independent methods:

\noindent$\bullet$ Euclidean quantum gravity [Hawking, unpublished];

\noindent$\bullet$ loop quantum gravity [Ashtekar \& Bojowald, unpublished];

\noindent$\bullet$ this qualitative classical analysis with minimal assumptions
about ``correct'' quantum gravity,

that a singularity never forms

and that the ingoing radiation has negative energy density during the entire
evaporation phase.

\bigskip\noindent\magenta{\em Orthodox paradox}\black

\noindent$\bullet$ Logically there are no true paradoxes, just
misunderstandings, usually simple but fundamental.

The \magenta{\em disinformation problem\black} for black holes: defining them
by event horizons as regions of no escape.

\noindent$\bullet$ Understanding black holes locally by trapping horizons, the
supposed paradox disappears.

\end{document}